\documentclass{ws-p9-75x6-50}

\begin{document}

\def\beq{\begin{equation}}
\def\eeq{\end{equation}}
\def\d{{\delta}}
\def\4G{{\delta G^{st}}}
\def\5G{{\delta {}^{(5)}}G}

\title{Cosmological evolution in a brane universe}

\author{David Langlois}

\address{ Institut d'Astrophysique de Paris, \\
Centre National de la Recherche Scientifique,\\
98bis Boulevard Arago, 75014 Paris, France \\
and\\
D\'epartement d'Astrophysique Relativiste et de Cosmologie,\\
Centre National de la Recherche Scientifique,\\
Observatoire de Paris, 92195 Meudon Cedex, France}


\maketitle

\abstracts{The idea of extra-dimensions has recently gone through a renewal with the 
hypothesis, suggested by recent developments in string theory, that ordinary
matter is confined to a sub-space, called brane, embedded in a higher 
dimensional spacetime. I summarize here some consequences in cosmology 
of this type of models. The most remarkable aspect is that the Friedmann 
laws, which govern the expansion of the Universe, are modified. An important 
direction of research is the study of cosmological perturbations and the 
possible signature of extra-dimensions in cosmological observations. 
 }

\section{Homogeneous brane cosmology}

It has been recently suggested that there could exist extra-dimensions
which are not accessible to  ordinary matter, in the sense that 
matter would be  confined to a three-dimensional subspace, 
or {\it brane}, within a higher dimensional space or {\it bulk}. 
In this context, a lot of attention has been devoted to cosmology, 
essentially to five-dimensional models where our Universe would be 
a hypersurface.

One of the first striking results was that, when solving 
the  five-dimensional Einstein's equations $G_{AB}\equiv R_{AB}-R g_{AB}/2=
\kappa^2 T_{AB}$,  the matter content of 
the brane appears {\it quadratically} \cite{bdl99}
 in Friedmann's  equations  instead of linearly as in standard cosmology.
In the case of an empty bulk, one would thus find a cosmological evolution 
incompatible with our understanding of nucleosynthesis.
 A way out has been 
found by applying the Randall-Sundrum idea \cite{rs99b} 
to cosmology \cite{cosmors,bdel99}, i.e. considering
an Anti-de Sitter  bulk spacetime (with 
a negative cosmological constant $\Lambda$) and a tension in the brane.  
The (assumed) cancellation of $\Lambda$ with the square of the brane tension 
$\sigma$
leads to the new Friedmann equations \cite{kraus,bdel99}
\beq
H^2= {8\pi G\over 3}\left(\rho+{\rho^2\over 2\sigma}\right),
\label{new_friedmann}
\eeq
where $H$ is the Hubble parameter in the brane, $\rho$ the cosmological energy
density in the brane. And Newton's constant is related to the brane tension 
by $8\pi G=\kappa^4\sigma/6$.
This equation gives the usual evolution in the
low energy regime $\rho\ll\sigma$ and quadratic corrections in the 
high energy regime $\rho > \sigma$.

\section{Cosmological perturbations in brane cosmology}

The next step is obviously to investigate what will be the influence of 
extra-dimensions on the {\it cosmological perturbations} and their evolution. 
Several pioneering works have developed formalisms to handle the cosmological 
perturbations for a brane-universe in a five-dimensional spacetime 
\cite{perturbations}.

The study of perturbations during a de Sitter phase 
 in the brane is made easier 
by the fact that the background evolution is rather simple.
One can for example compute explicitly the spectrum 
of gravitational waves generated   in a de Sitter phase on the 
brane \cite{lmw00}.

For radiation or matter dominated eras, the evolution of perturbations is much
more complicated. However, it is possible to rewrite the evolution equations
for the cosmological perturbations in the brane in a form very close to the 
equations of standard cosmology 
with two types of corrections: a. corrections due to the unconventional 
evolution of the homogeneous solution (in the high energy regime), 
which change the background-dependent 
coefficients of the equations; b. corrections due to the curvature along the 
fifth dimension, which act as source terms, i.e. `active seeds', 
 in the evolution equations \cite{l00B}. This reformulation of the 
five-dimensional equations makes transparent the way the perturbations due 
to the fifth dimension could have an impact, from the 
point of view of  a brane  observer,
for instance 
on the cosmic Microwave Background
anisotropies \cite{lmsw00}, but a quantitative 
analysis depends on the specific 
distribution of gravitational waves in the bulk.


\begin{thebibliography}{99}


\bibitem{bdl99}
P. Bin{\'e}truy, C. Deffayet, D. Langlois, {\it Nucl. Phys.} 
{\bf B 565}, 269 (2000) [hep-th/9905012].

\bibitem{rs99b}
L. Randall, R. Sundrum,  {\it Phys. Rev. Lett.} {\bf 83}, 4690
(1999) [hep-th/9906064].



\bibitem{cosmors}
C. Cs\'aki, M. Graesser, C. Kolda, J. Terning, {\it Phys. Lett.} {\bf B462}, 34 (1999)
;
J.M. Cline, C. Grojean, G. Servant,  {\it Phys. Rev. Lett.} {\bf 83}, 4245
  (1999).

\bibitem{bdel99} P. Bin{\'e}truy, C. Deffayet, U. Ellwanger, 
D. Langlois, {\it Phys. Lett.} {\bf B 477}, 285 (2000) [hep-th/9910219].


\bibitem{kraus} P. Kraus, JHEP 9912, 011 (1999) [hep-th/9910149]; 
T. Shiromizu, K. Maeda, M. Sasaki, {\it Phys. Rev.} {\bf D} 62, 024012 (2000)
 [gr-qc/9910076];
E. Flanagan,  S. Tye, I. Wasserman, 
{\it Phys. Rev.} {\bf D} 62, 044039 (2000) [hep-th/9910498]



\bibitem{perturbations} H. Kodama, A. Ishibashi, O. Seto, 
{\it Phys. Rev.} {\bf D} 62, 064022 (2000) [hep-th/0004160]; 
 R. Maartens, {\it Phys. Rev.} {\bf D} 62, 084023 (2000) 
  [hep-th/0004166];
D. Langlois, 
{\it Phys. Rev.} {\bf D} 62, 126012 (2000) [hep-th/0005025];
C. van de Bruck, M. Dorca, R. Brandenberger, A. Lukas, 
{\it Phys. Rev.} {\bf D} 62, 123515 (2000)
[hep-th/0005032]; K. Koyama, J. Soda, 
{\it Phys. Rev.} {\bf D} 62, 123502 (2000) [hep-th/0005239];
 N. Deruelle, T. Dolezel, J. Katz, hep-th/0010215. 


\bibitem{lmw00} D. Langlois, R. Maartens, D. Wands, {\it Phys. Lett.}
{\bf B} 489, 259 (2000
)
[hep-th/0006007] 

\bibitem{l00B} D. Langlois, ``Evolution of cosmological perturbations 
in a brane-universe'', hep-th/0010063, to appear in {\it Phys. Rev. Lett}.

\bibitem{lmsw00}  D. Langlois, R. Maartens, M. Sasaki, D. Wands,
``Large-scale cosmological perturbations on the brane'',  hep-th/0012044.



\end{thebibliography}
\end{document}